\documentclass[11pt, a4paper]{article}
\pdfoutput=1
\usepackage{jheppub}
\usepackage{amsmath}
\usepackage{amsfonts}
\usepackage{amssymb}
\usepackage{amsthm}
\usepackage{mathtools}
\usepackage{mathrsfs}
\usepackage[caption=false, singlelinecheck=false, justification=raggedright]{subfig}
\usepackage{txfonts}
\usepackage{xspace}
\usepackage{url}

\newcommand{\modulus}[1]{\left| #1 \right|}

\def\Dbar  {\kern 0.2em\overline{\kern -0.2em D}{}\xspace}
\def\Db    {\ensuremath{\Dbar}\xspace}
\def\Dzb   {\ensuremath{\overline{D^0}}\xspace}
\def\DzDzb {\ensuremath{D^0 {\kern -0.16em \Dzb}}\xspace}
\def\Dp    {\ensuremath{D^+}\xspace}

\def\Dm    {\ensuremath{D^-}\xspace}
\def\Dz    {\ensuremath{D^0}\xspace}

\def\Dzbs   {\ensuremath{\Dbar^{*0}}\xspace}
\def\DzsDzbs{\ensuremath{D^{*0} {\kern -0.16em \Dzbs}}\xspace}

\def\Bbar  {\kern 0.18em\overline{\kern -0.18em B}{}\xspace}

\def\Bzb   {\ensuremath{\Bbar^0}\xspace}
\def\BzBzb {\ensuremath{B^0 {\kern -0.16em -\Bzb}}\xspace}
\def\Bp    {\ensuremath{B^+}\xspace}
\def\Bpm   {\ensuremath{B^\pm}\xspace}
\def\Bm    {\ensuremath{B^-}\xspace}
\def\Bz    {\ensuremath{B^0}\xspace}

\def\Bbar  {\kern 0.18em\overline{\kern -0.18em B}{}\xspace}

\def\Bzb   {\ensuremath{\Bbar^0}\xspace}
\def\BzBzb {\ensuremath{B^0 {\kern -0.16em -\Bzb}}\xspace}

\def\PA   {\ensuremath{P\!A}\xspace}
\def\PE   {\ensuremath{P\!E}\xspace}

\def\nn {\nonumber}

\def\Apm   {A_{+-}}

\def\Azm   {A_{0-}}
\def\Abpm  {\overline{A}_{+-}}
\def\Abpz  {\overline{A}_{+0}}
\def\Abzm  {\overline{A}_{0-}}
\def\Azz   {A_{00}}
\def\Abzz  {\overline{A}_{00}}
\def\Bpm   {B_{+-}}
\def\Bzz   {B_{00}}
\def\Bch   {B_{\text{ch}}}

\def\BBpm   {\mathscr{B}_{+-}}
\def\BBzz   {\mathscr{B}_{00}}
\def\BBch   {\mathscr{B}_{\text{ch}}}
\def\Cpm   {C_{+-}}
\def\Czz   {C_{00}}
\def\ACP   {A_{\text{CP}}}

\def\CP{\textsl{CP}\xspace}

\allowdisplaybreaks

\begin{document}
\title{Prediction of the \textsl{CP} asymmetry \texorpdfstring{$C_{00}$}{C00} in
	\texorpdfstring{$B^0 \to D^0\overline{D^0}$}{B0 -> D0 D0-bar} decay} %

\author[a]{Dibyakrupa Sahoo,} %
\author[b]{Hai-Yang Cheng,} %
\author[b,c,d]{Cheng-Wei Chiang,} %
\author[a]{C.~S.~Kim} %
\author[e]{and Rahul Sinha} %

\affiliation[a]{Department of Physics and IPAP, Yonsei University, Seoul 120-749, Korea}

\affiliation[b]{Institute of Physics, Academia Sinica, Taipei, Taiwan 11529, Republic of China} %

\affiliation[c]{Department of Physics, National Taiwan University, Taipei, Taiwan 10617, Republic of China} %
\affiliation[d]{Kavli IPMU, University of Tokyo, Kashiwa, 277-8583, Japan}

\affiliation[e]{The Institute of Mathematical Sciences, Taramani, Chennai 600113, India} %

\emailAdd{sahoodibya@yonsei.ac.kr} %
\emailAdd{phcheng@phys.sinica.edu.tw} %
\emailAdd{chengwei@phys.ntu.edu.tw} %
\emailAdd{cskim@yonsei.ac.kr} %
\emailAdd{sinha@imsc.res.in} %

\date{\today}

\abstract{Of all $B \to D \overline{D}$ decays, the $B^0 \to D^0 \overline{D^0}$
	decay has the smallest observed branching ratio as it takes place primarily via
	the suppressed $W$-exchange diagram. The \textsl{CP} asymmetry for this mode is
	yet to be measured experimentally. By exploiting the relationship among the
	decay amplitudes of $B \to D\overline{D}$ decays (using isospin and topological
	amplitudes) we are able to relate the \textsl{CP} asymmetries and branching
	ratios by a simple expression. This enables us to predict the \textsl{CP}
	asymmetry $C_{00}$ in $B^0 \to D^0 \overline{D^0}$. While the predicted central
	values of $C_{00}$ are outside the physically allowed region, they are currently
	associated with large uncertainties owing to the large errors in the
	measurements of the $B^0 \to D^0 \overline{D^0}$ branching ratio ($B_{00}$), the
	other \textsl{CP} asymmetries $C_{+-}$ (of $B^0 \to D^+ D^-$) and
	$A_{\text{CP}}$ (of $B^+ \to D^+ \overline{D^0}$). With a precise determination
	of $B_{00}$, $C_{+-}$ and $A_{\text{CP}}$, one can use our analytical result to
	predict $C_{00}$ with a reduced error and compare it with the experimental
	measurement when it becomes available. The correlation between $B_{00}$ and
	$C_{00}$ is an interesting aspect that can be probed in ongoing and future
	particle physics experiments such as LHCb and Belle~II.}

\arxivnumber{1709.08301}

\maketitle

\section{Introduction}\label{sec:introduction}

It is very well known that violation of \CP symmetry, the combined symmetry of
charge conjugation (\textsl{C}) and parity (\textsl{P}), is essential for the
matter-antimatter asymmetry observed in our Universe \cite{Sakharov:1967dj}. All
observed \CP violation in $K$ and $B$ meson decays are successfully explained by
the Cabibbo-Kobayashi-Maskawa (CKM) matrix
\cite{Cabibbo:1963yz,Kobayashi:1973fv} which is a cornerstone of the standard
model (SM) of particle physics. However, \CP violation as we know in the SM is
not sufficient to account for the observed baryon asymmetry in our Universe
\cite{Gavela:1993ts,Gavela:1994dt,Huet:1994jb}. Therefore, experimental searches
are still going on to find out possibly new sources of \CP violation beyond the
SM. In this context, study of decays of heavy flavor mesons, especially the $B$
mesons, has played an important role (see Refs.~\cite{Antonelli:2009ws,
	Hocker:2006xb, Artuso:2015swg, Gershon:2016fda} for some recent reviews). In
this work we shall analyze the $B \to D \overline{D}$ decays, in particular $\Bp
\to \Dp \Dzb$, $\Bz \to \Dz\Dzb$, $\Bz \to \Dp\Dm$ and their \CP conjugate
processes, with a view to predict the \CP-violating parameter $C_{00}$ for $B^0
\to D^0 \overline{D^0}$, which has not yet been measured experimentally.

In our analysis we shall exploit the isospin symmetry, which is known to be a
very useful symmetry in the study of various hadronic decays, most notably in
many $B$ meson decays. The existing literature is replete with many interesting
studies of double charm decays of the $B$ mesons~\cite{Bel:2015wha,
	Jung:2014jfa, Mohammadi:2011zz, Lu:2010gg, Li:2009xf, Kim:2008ex, Gronau:2008ed,
	Li:2007rk, Fleischer:2007zn, Chen:2005rp, Datta:2003va, Xing:1999yx,
	Pham:1999fy, Fleischer:1999nz, Xing:1998ca, Sanda:1997pm, Gronau:1995hm,
	Kramer:1994in, Aleksan:1993qk}. Some of these works~\cite{Jung:2014jfa,
	Gronau:2008ed, Xing:1999yx, Sanda:1997pm, Gronau:1995hm} also analyze the $B \to
D \Dbar$ decays in terms of isospin symmetry. The $B \to D\Db$ decays get
contributions from currents that change isospin by $1/2$ and $3/2$. Without
making any assumptions regarding the sizes of these contributions, we find out
their upper and lower limits. Another useful method to study various hadronic
decays of the $B$ and $D$ mesons is the topological diagram
approach~\cite{Zeppenfeld:1980ex, Chau:1982da, Chau:1986du, Chau:1987tk,
	Chau:1990ay, Kim:1998sh, Chiang:2003jn, Chiang:2003rb, Chiang:2003pm,
	Chiang:2004nm, Chiang:2006ih, Chiang:2007qh, Chiang:2008vc, Cheng:2010ry,
	Cheng:2010vk, Cheng:2011qh, Cheng:2012wr, Cheng:2012xb, Cheng:2016ejf}. We have
used this approach in conjunction with the isospin symmetry to derive a simple
expression relating all the branching ratios and \CP asymmetries under
consideration. This enables us to predict the \CP asymmetry $\Czz$ of the $\Bz
\to \DzDzb$ mode.

This paper is structured as follows. In Sec.~\ref{sec:isospin}, we do the
isospin decomposition of the concerned decay amplitudes, keeping both $\Delta I
=1/2$ and $3/2$ contributions, and provide the upper and lower limits on them.
The decay amplitudes are then analyzed from the perspective of topological
amplitudes in Sec.~\ref{sec:topological-diagrams}. This leads to an expression
for $\Czz$ in Sec.~\ref{sec:C00-expression}. It is followed by a relevant
numerical analysis in Sec.~\ref{sec:numerical}, showing how precision
measurements can improve our predictions in the future. Finally we conclude in
Sec.~\ref{sec:conclusion}, highlighting the important results of our analysis.

\section{Isospin analysis of \texorpdfstring{$B \to D \overline{D}$}{B -> D Dbar} decay amplitudes}\label{sec:isospin}

\subsection{Isospin decomposition of the decay amplitudes}

In the $B \to D\Db$ decays, the initial $B$ meson has isospin $I = \frac{1}{2}$
and the final $D\Db$ state has isospin $I=0,1$. The effective weak interaction
Hamiltonian driving these decays has currents which change the isospin by $1/2$
and $3/2$. Thus the decay amplitudes for the various $B \to D\Db$ decays can be
decomposed under isospin consideration as follows:
\begin{subequations}\label{eq:Amp-iso}
\begin{align}
\Apm & \equiv \textrm{A}(\Bzb \to \Dp\Dm) = \frac{1}{\sqrt{2}} \left( A_1 +
B_1 + A_0 \right),\\%
\Azz & \equiv \textrm{A}(\Bzb \to \Dz\Dzb) = \frac{1}{\sqrt{2}} \left( A_1 +
B_1 - A_0 \right),\\%
\Azm & \equiv \textrm{A}(\Bm \to \Dz\Dm) = \frac{1}{\sqrt{2}} \left( 2 A_1 -
B_1 \right),%
\end{align}
\end{subequations}
where $A_0$ and $A_1$ are the isospin amplitudes facilitated by $\Delta I=1/2$
current to the isospin $I=0,1$ final states respectively and $B_1$ denotes the
isospin amplitude with $\Delta I=3/2$ current to $I=1$ final state. The
conjugate amplitudes are defined as:
\begin{subequations}\label{eq:Ampb-iso}
\begin{align}
\Abpm & \equiv \textrm{A}(\Bz \to \Dp\Dm) = \frac{1}{\sqrt{2}} \left(
\overline{A}_1 + \overline{B}_1 + \overline{A}_0 \right),\\ %
\Abzz & \equiv \textrm{A}(\Bz \to \Dz\Dzb) = \frac{1}{\sqrt{2}} \left(
\overline{A}_1 + \overline{B}_1 - \overline{A}_0 \right),\\ %
\Abpz & \equiv \textrm{A}(\Bp \to \Dp\Dzb) = \frac{1}{\sqrt{2}} \left( 2
\overline{A}_1 - \overline{B}_1 \right).
\end{align}
\end{subequations}
It is easy to notice from Eqs.~\eqref{eq:Amp-iso} and \eqref{eq:Ampb-iso} that
the amplitudes satisfy the following relations:
\begin{subequations}
\begin{align}
\Apm - \Azz &= \sqrt{2}~A_0,\\%
\Abpm - \Abzz &= \sqrt{2}~\overline{A}_0,\\%
\Apm + \Azz + 2 \Azm &= 3 \sqrt{2}~A_1,\\%
\Abpm + \Abzz + 2 \Abzm &= 3 \sqrt{2}~\overline{A}_1,\\%
\Apm + \Azz - \Azm &= \frac{3}{\sqrt{2}} B_1, \label{eq:quadrilateral1}\\ %
\Abpm + \Abzz - \Abpz &= \frac{3}{\sqrt{2}}
\overline{B}_1.\label{eq:quadrilateral2}
\end{align}
\end{subequations}
If one were to neglect $B_1$ altogether, one would get the relations
$\Apm+\Azz=\Azm$ and $\Abpm+\Abzz=\Abpz$ as given by Sanda and
Xing~\cite{Sanda:1997pm}.

\begin{figure}[ht]
\centering%
\includegraphics[scale=1]{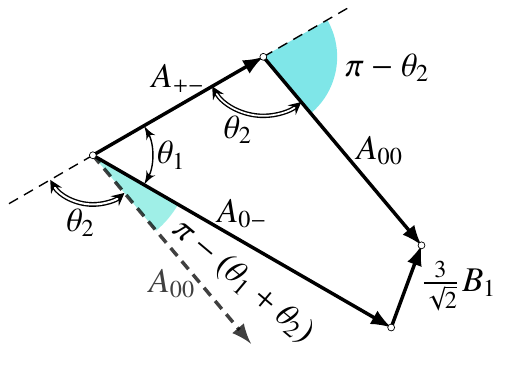}%
\caption{A representative quadrilateral formed by $\Apm$, $\Azz$, $\Azm$ and
	$\frac{3}{\sqrt{2}} B_1$, depicting Eq.~\eqref{eq:quadrilateral1} graphically
	(not drawn to scale). Another quadrilateral can be drawn similarly for the
	conjugate amplitudes, for which we denote the analogous angles by $\theta'_1$,
	$\theta'_2$. The angles $\theta'_1$ and $\theta'_2$ can generally be different
	from $\theta_1$ and $\theta_2$.}%
\label{fig:quadrilateral-1}
\end{figure}

Since amplitudes are complex quantities, they are denoted by vectors in the
complex plane. The amplitudes $\Apm$, $\Azz$ and $\Azm$ along with
$\frac{3}{\sqrt{2}} B_1$ (see Eq.~\eqref{eq:quadrilateral1}) form a
quadrilateral as shown in Fig.~\ref{fig:quadrilateral-1}. Depending on the
values for angles $\theta_1$ and $\theta_2$, the quadrilateral can be either a
\textit{simple} quadrilateral or a \textit{self-intersecting} quadrilateral.
Once again, if $B_1 =0$, we would get back to the triangles of Sanda and
Xing~\cite{Sanda:1997pm}. It is, therefore, interesting to find out how large
the magnitudes of $B_1$ and $\overline{B}_1$ can be per the current experimental
observations. Before we get into finding out the limits on $\modulus{B_1}$ and
$\modulus{\overline{B}_1}$, let us first write down the expressions for the
experimental observables.

\subsection{The experimental observables}

The experimental observables we shall use in our analysis are the branching
ratios\footnote{From dimensional analysis of the expressions for branching
	ratios in Eq.~\eqref{eq:observables} it is easy to see that the amplitudes have
	mass-dimension 1 in our case.} and \CP asymmetries which are defined as,
\begin{subequations}\label{eq:observables}
\begin{align}
\Bpm &= \frac{1}{2} \tau_0 \frac{\sqrt{\lambda\left(m_{\Bz}^2, m_{\Dp}^2,
m_{\Dp}^2 \right)}}{16 \pi \, m_{\Bz}^3} \left( \modulus{\Apm}^2 +
\modulus{\Abpm}^2 \right),\\ %
\Bzz &= \frac{1}{2} \tau_0 \frac{\sqrt{\lambda\left(m_{\Bz}^2, m_{\Dz}^2,
m_{\Dz}^2 \right)}}{16 \pi \, m_{\Bz}^3} \left( \modulus{\Azz}^2 +
\modulus{\Abzz}^2 \right),\\ %
\Bch &= \frac{1}{2} \tau_+ \frac{\sqrt{\lambda\left(m_{\Bp}^2, m_{\Dp}^2,
m_{\Dz}^2 \right)}}{16 \pi \, m_{\Bp}^3} \left( \modulus{\Azm}^2 +
\modulus{\Abpz}^2 \right),\\ %
\Cpm &= \frac{\modulus{\Apm}^2 - \modulus{\Abpm}^2}{\modulus{\Apm}^2 +
\modulus{\Abpm}^2},\\ %
\Czz &= \frac{\modulus{\Azz}^2 - \modulus{\Abzz}^2}{\modulus{\Azz}^2 +
\modulus{\Abzz}^2},\\ %
\ACP &= \frac{\modulus{\Azm}^2 - \modulus{\Abpz}^2}{\modulus{\Azm}^2 +
\modulus{\Abpz}^2},
\end{align}
\end{subequations}
where $m_{i}$ represents the mass of particle $i$, $\tau_0$, $\tau_+$ are
the mean lifetimes of $\Bz$, $\Bp$, respectively, and
\begin{align*}
\lambda(x,y,z) = x^2 + y^2 + z^2 - 2xy - 2yz - 2zx ~.
\end{align*}
The subscript in $\Bch$
denotes the fact that we are dealing with the decay of a charged $B$ meson in
this case.

For simplicity we shall define the `scaled' branching ratios (which have mass
dimension 2) and express the \CP asymmetries in terms of them as shown below:
\begin{subequations}\label{eq:modified-observables}
\begin{align}
\BBpm &= \frac{16 \pi m_{\Bz}^3}{\tau_0 \sqrt{\lambda\left( m_{\Bz}^2,
		m_{\Dp}^2, m_{\Dp}^2 \right)}} \Bpm = \frac{1}{2} \left( \modulus{\Apm}^2 +
\modulus{\Abpm}^2 \right),\\ %
\BBzz &= \frac{16 \pi m_{\Bz}^3}{\tau_0 \sqrt{\lambda\left( m_{\Bz}^2,
		m_{\Dz}^2, m_{\Dz}^2 \right)}} \Bzz = \frac{1}{2} \left( \modulus{\Azz}^2 +
\modulus{\Abzz}^2 \right),\\ %
\BBch &= \frac{16 \pi m_{\Bp}^3}{\tau_+ \sqrt{\lambda\left( m_{\Bp}^2,
		m_{\Dp}^2, m_{\Dz}^2 \right)}} \Bch = \frac{1}{2} \left( \modulus{\Azm}^2 +
\modulus{\Abpz}^2 \right),\\ %
\Cpm &= \frac{1}{2\BBpm} \left( \modulus{\Apm}^2 - \modulus{\Abpm}^2 \right),\\
\Czz &= \frac{1}{2\BBzz} \left( \modulus{\Azz}^2 - \modulus{\Abzz}^2 \right),\\
\ACP &= \frac{1}{2\BBch} \left( \modulus{\Azm}^2 - \modulus{\Abpz}^2 \right).
\end{align}
\end{subequations}

From the definitions of the observables given in
Eq.~\eqref{eq:modified-observables}, we can easily obtain the following
relations:
\begin{subequations}\label{eq:amp-mods}
\begin{align}
\modulus{\Apm}  &= \sqrt{\BBpm \left( 1+ \Cpm  \right)}~,\\ %
\modulus{\Azz}  &= \sqrt{\BBzz \left( 1+ \Czz  \right)}~,\\ %
\modulus{\Azm}  &= \sqrt{\BBch \left( 1+ \ACP  \right)}~,\\ %
\modulus{\Abpm} &= \sqrt{\BBpm \left( 1- \Cpm  \right)}~,\\ %
\modulus{\Abzz} &= \sqrt{\BBzz \left( 1- \Czz  \right)}~,\\ %
\modulus{\Abpz} &= \sqrt{\BBch \left( 1- \ACP  \right)}~.
\end{align}
\end{subequations}
Since the \CP asymmetries must always lie between $-1$ and $1$, {\it i.e.},\ $-1
\leqslant \Czz,\Cpm,\ACP \leqslant 1$, the moduli of the amplitudes are always
ensured to be positive and real by definition.

\subsection{Upper and lower limits on the magnitudes of isospin amplitudes}

From Fig.~\ref{fig:quadrilateral-1} it is easy to show that
\begin{align}
\modulus{B_1}^2 &= \frac{2}{9} \Big( \modulus{\Apm}^2 + \modulus{\Azz}^2 +
\modulus{\Azm}^2 - 2 \modulus{\Apm} \modulus{\Azz} \cos\theta_2 \nn\\%
&~+ 2 \modulus{\Azz} \modulus{\Azm} \cos\left( \theta_1 + \theta_2 \right) - 2
\modulus{\Azm} \modulus{\Apm} \cos\theta_1 \Big).\label{eq:B1exp}
\end{align}
Considering the conjugate amplitudes, we would get
\begin{align}
\modulus{\overline{B}_1}^2 &= \frac{2}{9} \Big( \modulus{\Abpm}^2 +
\modulus{\Abzz}^2 + \modulus{\Abpz}^2 - 2 \modulus{\Abpm} \modulus{\Abzz}
\cos\theta'_2 \nn\\*%
&~+ 2 \modulus{\Abzz} \modulus{\Abpz} \cos\left( \theta'_1 + \theta'_2 \right) -
2 \modulus{\Abpz} \modulus{\Abpm} \cos\theta'_1 \Big),\label{eq:B1bexp}
\end{align}
where the angles $\theta'_1$, $\theta'_2$ denote the fact that they are
necessarily different from the analogous angles $\theta_1$, $\theta_2$. Now the
limits on $\modulus{B_1}$ \bigg(and $\modulus{\overline{B}_1}$\bigg) can be
obtained by taking some specific values for the angles $\theta_1$ and $\theta_2$
of Fig.~\ref{fig:quadrilateral-1} (or $\theta'_1$ and $\theta'_2$), as well as
for the moduli of the decay amplitudes, as shown below.

\paragraph{Maximum} %
The maximum value for $\modulus{B_1}$ is obtained when $\Apm$, $\Azz$ and
$-\Azm$ are all directed along the same direction, {\it i.e.}, when the angles in
Fig.~\ref{fig:quadrilateral-1} are set to the values $\theta_1 = \pi =
\theta_2$,
\begin{equation}\label{eq:bound-max-1}
\modulus{B_1}^2_{\textrm{max}} = \frac{2}{9}\Big( \modulus{\Apm} +
\modulus{\Azz} + \modulus{\Azm} \Big)^2.
\end{equation}
We can consider all the conjugate amplitudes in a similar manner, and this would
lead to the maximum for $\modulus{\overline{B}_1}$,
\begin{equation}\label{eq:bound-max-2}
\modulus{\overline{B}_1}^2_{\textrm{max}} = \frac{2}{9} \Big( \modulus{\Abpm} +
\modulus{\Abzz} + \modulus{\Abpz} \Big)^2.
\end{equation}

\paragraph{Minimum} %
There are two interesting scenarios for finding the minimum of
$\modulus{B_1}^2$. In the first case, the quadrilateral is squashed into a
straight line (analogous to the situation for maximum), while in the later case,
for $\modulus{B_1}_{\text{min}}=0$, the quadrilateral is transformed into a
triangle. These two cases can be easily distinguished from each other by first
arranging the three amplitude moduli in either increasing or decreasing order.
Then we take the sum of the two smaller moduli. If the largest modulus is
greater than the sum of the two smaller moduli, then the amplitudes can never
form a triangle, i.e.\ $\modulus{B_1}_{\textrm{min}} \neq 0$. In the case where
the largest modulus is smaller than the sum of the two smaller moduli, the
amplitudes can form a triangle resulting in $\modulus{B_1}_{\textrm{min}}=0$.
When $\modulus{B_1}_{\textrm{min}} \neq 0$ we have the following three
possibilities,
\begin{equation}\label{eq:bound-min-1}
\modulus{B_1}^2_{\textrm{min}} =
\begin{cases}
\frac{2}{9} \Big( - \modulus{\Apm} + \modulus{\Azz} + \modulus{\Azm} \Big)^2
&\textrm{for } \theta_1 = 0 = \theta_2,\\%
\frac{2}{9} \Big( \modulus{\Apm} + \modulus{\Azz} - \modulus{\Azm} \Big)^2
&\textrm{for } \theta_1 = 0, \theta_2 = \pi,\\%
\frac{2}{9} \Big( \modulus{\Apm} - \modulus{\Azz} + \modulus{\Azm} \Big)^2
&\textrm{for } \theta_1 = \pi, \theta_2 = 0.
\end{cases}
\end{equation}
Once again, considering the conjugate amplitudes would give the set of three
minima for $\modulus{\overline{B}_1}$,
\begin{equation}\label{eq:bound-min-2}
\modulus{\overline{B}_1}^2_{\textrm{min}} =
\begin{cases}
\frac{2}{9} \Big( - \modulus{\Abpm} + \modulus{\Abzz} + \modulus{\Abpz} \Big)^2
&\textrm{for } \theta'_1 = 0 = \theta'_2,\\%
\frac{2}{9} \Big( \modulus{\Abpm} + \modulus{\Abzz} - \modulus{\Abpz} \Big)^2
&\textrm{for } \theta'_1 = 0, \theta'_2 = \pi,\\%
\frac{2}{9} \Big( \modulus{\Abpm} - \modulus{\Abzz} + \modulus{\Abpz} \Big)^2
&\textrm{for } \theta'_1 = \pi, \theta'_2 = 0.
\end{cases}
\end{equation}

It is important to note that following steps similar to the ones presented here,
it is also possible to give expressions for maximum and minimum of
$\modulus{A_1}$ and $\modulus{\overline{A}_1}$ as follows,
\begin{subequations}\label{eq:A1-limits}
\begin{align}
\modulus{A_1}^2_{\textrm{max}} &= \frac{1}{18}\Big( \modulus{\Apm} +
\modulus{\Azz} + 2\modulus{\Azm} \Big)^2, \\%
\modulus{\overline{A}_1}^2_{\textrm{max}} &= \frac{1}{18} \Big( \modulus{\Abpm}
+ \modulus{\Abzz} + 2\modulus{\Abpz} \Big)^2, \\%
\modulus{A_1}^2_{\textrm{min}} &=
\begin{cases}
\frac{1}{18} \Big( - \modulus{\Apm} + \modulus{\Azz} + 2\modulus{\Azm}
\Big)^2,\\%
\frac{1}{18} \Big( \modulus{\Apm} + \modulus{\Azz} - 2\modulus{\Azm} \Big)^2,\\%
\frac{1}{18} \Big( \modulus{\Apm} - \modulus{\Azz} + 2\modulus{\Azm} \Big)^2,
\end{cases} \\%
\modulus{\overline{A}_1}^2_{\textrm{min}} &=
\begin{cases}
\frac{1}{18} \Big( - \modulus{\Abpm} + \modulus{\Abzz} + 2\modulus{\Abpz}
\Big)^2,\\%
\frac{1}{18} \Big( \modulus{\Abpm} + \modulus{\Abzz} - 2\modulus{\Abpz}
\Big)^2,\\%
\frac{1}{18} \Big( \modulus{\Abpm} - \modulus{\Abzz} + 2\modulus{\Abpz} \Big)^2.
\end{cases}
\end{align}
\end{subequations}
Furthermore, it is trivial to find out the maximum and minimum of
$\modulus{A_0}$ and $\modulus{\overline{A}_0}$,
\begin{subequations}\label{eq:A0-limits}
\begin{align}
\modulus{A_0}^2_{\text{max}} &= \frac{1}{2} \Big( \modulus{\Apm} +
\modulus{\Azz} \Big)^2,\\%
\modulus{A_0}^2_{\text{min}} &= \frac{1}{2} \Big( \modulus{\Apm} -
\modulus{\Azz} \Big)^2,\\%
\modulus{\overline{A}_0}^2_{\text{max}} &= \frac{1}{2} \Big( \modulus{\Abpm} +
\modulus{\Abzz} \Big)^2,\\%
\modulus{\overline{A}_0}^2_{\text{min}} &= \frac{1}{2} \Big( \modulus{\Abpm} -
\modulus{\Abzz} \Big)^2.
\end{align}
\end{subequations}
We shall provide a numerical comparison of the allowed maximum and minimum
values for $\modulus{A_0}$, $\modulus{A_1}$, $\modulus{B_1}$,
$\modulus{\overline{A}_0}$, $\modulus{\overline{A}_1}$,
$\modulus{\overline{B}_1}$ in Sec.~\ref{sec:numerical}.

Thus far, we have considered the two possible quadrilaterals separately.
However, the amplitudes and their \CP conjugate amplitudes are related to one
another via the strong and weak phases. In order to do an analysis keeping both
strong and weak phases into account, we shall consider the various quark
diagrams (also known as topological diagrams) contributing to the $B \to D\Db$
decays under our consideration.

\section{Analysis of \texorpdfstring{$B \to D \overline{D}$ }{B -> D Dbar} decay amplitudes under diagrammatic approach}\label{sec:topological-diagrams}

\subsection{Contributing topological diagrams}

The $B \to D\Db$ decays are facilitated by various topological diagrams, which are
enunciated below.
\begin{enumerate}
	\item The $\Bp \to \Dp \Dzb$ decay gets contributions from color-allowed tree,
	$W$-annihilation, QCD-penguin, QCD-penguin exchange, color-suppressed
	electroweak-penguin and electroweak-penguin exchange diagrams.%
	\item The $\Bz \to \Dp \Dm$ decay gets contributions from color-allowed tree,
	$W$-exchange, QCD-penguin, QCD-penguin exchange, QCD-penguin annihilation,
	color-suppressed electroweak-penguin, electroweak-penguin exchange and
	electroweak-penguin annihilation diagrams.%
	\item The $\Bz \to \Dz\Dzb$ decay gets contributions from $W$-exchange,
	QCD-penguin annihilation and electroweak-penguin annihilation diagrams.
\end{enumerate}
Therefore, theoretically the branching ratio for $\Bz \to \Dz \Dzb$ is expected
to be smaller than those for $\Bp \to \Dp \Dzb$ and $\Bz \to \Dp \Dm$
\cite{Li:2007rk}. Contribution of each topological diagram is denoted by an
amplitude, the topological amplitude, multiplied by appropriate CKM matrix
elements. All $B \to D\Db$ decay amplitudes under our consideration are
proportional to $V_{Ub}^* V_{Ud}$ where $U$ can be $u,c,t$ and $V$ denotes the
CKM matrix. The relevant CKM unitarity condition for $B \to D\Db$ decays is
\begin{equation}\label{eq:unitarity_condition}
V^*_{ub} V_{ud} + V^*_{cb} V_{cd} + V^*_{tb} V_{td} = 0.
\end{equation}
In our subsequent discussions, we shall use the weak phase $\beta$ which is an angle of
the unitarity triangle associated with Eq.~\eqref{eq:unitarity_condition} and defined
as
\begin{equation}
\beta = \text{arg} \left( -\frac{V^*_{cb} V_{cd}}{V^*_{tb} V_{td}} \right).
\end{equation}

\subsection{Decomposition of decay amplitudes in terms of topological diagrams}

If we break up our decay amplitudes by using the topological amplitudes with
relevant CKM matrix elements, then we can relate the isospin amplitudes with
combinations of topological amplitudes. In general, we can use the unitarity
condition and the definition of weak phase $\beta$ to write down the decay
amplitudes as follows,
\begin{subequations}\label{eq:decay-amp-decomp}
\begin{align}
\Apm &= A'_{+-} + A''_{+-}~ e^{i\beta},\\%
\Azz &= A'_{00} + A''_{00}~ e^{i\beta},\\%
\Azm &= A'_{0-} + A''_{0-}~ e^{i\beta},
\end{align}
\end{subequations}
where
\begin{subequations}\label{eq:decay-amp-parts}
\begin{align}
A'_{+-} &= \frac{1}{\sqrt{2}} \left( \modulus{A'_1} e^{i\delta_1} +
\modulus{A'_0} e^{i\delta_0} + \modulus{B'_1} \right),\\%
A''_{+-} &= \frac{1}{\sqrt{2}} \left( \modulus{A''_1} e^{i\delta_1} +
\modulus{A''_0} e^{i\delta_0} + \modulus{B''_1} \right),\\%
A'_{00} &= \frac{1}{\sqrt{2}} \left( \modulus{A'_1} e^{i\delta_1} -
\modulus{A'_0} e^{i\delta_0} + \modulus{B'_1} \right),\\%
A''_{00} &= \frac{1}{\sqrt{2}} \left( \modulus{A''_1} e^{i\delta_1} -
\modulus{A''_0} e^{i\delta_0} + \modulus{B''_1} \right),\\%
A'_{0-} &= \frac{1}{\sqrt{2}} \left( 2\modulus{A'_1} e^{i\delta_1} -
\modulus{B'_1} \right),\\%
A''_{0-} &= \frac{1}{\sqrt{2}} \left( 2\modulus{A''_1} e^{i\delta_1} -
\modulus{B''_1} \right),
\end{align}
\end{subequations}
with $A'_0$, $A''_0$, $A'_1$, $A''_1$, $B'_1$, $B''_1$ being various components
of the isospin amplitudes, each with a distinct decomposition in terms of the
contributing topological amplitudes, and $\delta_1$, $\delta_0$ being the strong
phases measured with respect to $B_1$. The decomposition of various components
of the isospin amplitudes of Eq.~\eqref{eq:decay-amp-parts} in terms of the
topological amplitudes are as follows:
\begin{subequations}\label{eq:topological-decomposition}
\begin{align}
A'_0 &= \frac{1}{\sqrt{2}} \left( - T_c + E_c \right),\\%
A''_0 &= \frac{1}{3\sqrt{2}} \Big( 3E_t + 3P + 2 P^C_{EW} + 3\PE -
\PE_{EW} - 12 \PA -5 \PA_{EW} \Big),\\%
A'_1 &= \frac{1}{3\sqrt{2}} \left( -3T_c + E_c + 2 A_c \right),\\%
A''_1 &= \frac{1}{3\sqrt{2}} \Big( -E_t - 2 A_t + 3 P + 2 P^C_{EW} + 3
\PE + \PE_{EW} + \PA_{EW} \Big),\\%
B'_1 &= \frac{\sqrt{2}}{3} \left( E_c - A_c \right),\\%
B''_1 &= \frac{\sqrt{2}}{3} \left( - E_t + A_t + \PA_{EW} - \PE_{EW}
\right),
\end{align}
\end{subequations}
where the tree $T$, $W$-annihilation $A$, $W$-exchange $E$ are the topologies
with no quark loop, and QCD-penguin $P$, QCD-penguin annihilation $\PA$,
QCD-penguin exchange $\PE$, color-suppressed electroweak-penguin $P^C_{EW}$,
electroweak-penguin annihilation $\PA_{EW}$ and electroweak-penguin exchange
$\PE_{EW}$ are the dominant one loop topologies with top quark in the loop. Note
that for the no-loop topologies $N \in \{T,E,A\}$ the subscript in
Eq.~\eqref{eq:topological-decomposition} has the meaning that $N_x = N V_{xb}^*
V_{xd}$, and in the one loop topologies $L \in \{P,\PA,\PE,P^C_{EW},\PA_{EW},
\PE_{EW} \}$ in Eq.~\eqref{eq:topological-decomposition} the factor $V_{tb}^*
V_{td}$ is implicitly present. The up and charm quark contributions to one loop
topologies can also be considered in a similar manner. However, they do not
affect our analysis. Finally, the decay amplitudes for the conjugate processes
are obtained by switching the sign of the weak phase $\beta$ in
Eq.~\eqref{eq:decay-amp-decomp},
\begin{subequations}\label{eq:conju-decay-amp-decomp}
\begin{align}
\Abpm &= A'_{+-} + A''_{+-} e^{-i\beta},\\%
\Abzz &= A'_{00} + A''_{00} e^{-i\beta},\\%
\Abzm &= A'_{0-} + A''_{0-} e^{-i\beta}.
\end{align}
\end{subequations}
We shall now look at the consequences of the two amplitude decompositions we
have carried out.

\section{Consequence of decomposition of amplitudes using isospin and topological diagrams}\label{sec:C00-expression}

Starting from Eqs.~\eqref{eq:decay-amp-decomp}, \eqref{eq:decay-amp-parts} and
\eqref{eq:conju-decay-amp-decomp} one can write down the observables $\BBpm$,
$\Cpm$, $\BBzz$, $\Czz$, $\BBch$ and $\ACP$ in terms of the various isospin
amplitudes, strong and weak phases. The expressions for the \CP asymmetries are
given by
\begin{subequations}
\begin{align}
\Cpm &= \frac{\sin\beta}{\BBpm}\Bigg( -\left(\modulus{A''_0} \modulus{B'_1} -
\modulus{A'_0} \modulus{B''_1}\right)\sin\delta_0 - \left(\modulus{A''_1}
\modulus{B'_1} - \modulus{A'_1} \modulus{B''_1}\right)\sin\delta_1 \nonumber\\*%
& \hspace{6cm} - \left(\modulus{A''_0} \modulus{A'_1} - \modulus{A'_0}
\modulus{A''_1}\right)\sin\left( \delta_0 -\delta_1 \right) \Bigg),\\ %
\Czz &= \frac{\sin\beta}{\BBzz}\Bigg( \left(\modulus{A''_0} \modulus{B'_1} -
\modulus{A'_0} \modulus{B''_1}\right)\sin\delta_0 - \left(\modulus{A''_1}
\modulus{B'_1} - \modulus{A'_1} \modulus{B''_1}\right)\sin\delta_1 \nonumber\\*%
& \hspace{6cm} + \left(\modulus{A''_0} \modulus{A'_1} - \modulus{A'_0}
\modulus{A''_1}\right)\sin\left( \delta_0 -\delta_1 \right) \Bigg),\\ %
\ACP &= \frac{2\sin\beta}{\BBch}\left( \modulus{A''_1} \modulus{B'_1} -
\modulus{A'_1} \modulus{B''_1} \right)~\sin\delta_1.
\end{align}
\end{subequations}
With these results, it is straightforward to obtain the following important
expression which relates all the known branching ratios and \CP asymmetries for
the $B \to D\Db$ decays:
\begin{equation}
\BBpm~\Cpm + \BBzz~\Czz + \BBch~\ACP = 0.
\end{equation}
This simple relation can be used to predict $\Czz$ which is currently not
measured experimentally,
\begin{align}
\Czz &= -\frac{1}{\BBzz} \bigg( \BBpm~\Cpm + \BBch~\ACP \bigg)\nonumber\\%
&= - \frac{\sqrt{\lambda\left(m_{\Bz}^2, m_{\Dz}^2, m_{\Dz}^2 \right)}}{\Bzz}
\left( \frac{\Bpm \Cpm}{\displaystyle \sqrt{\lambda\left(m_{\Bz}^2, m_{\Dp}^2,
		m_{\Dp}^2 \right)}} + \frac{\Bch \ACP}{\displaystyle
	\sqrt{\lambda\left(m_{\Bp}^2, m_{\Dp}^2, m_{\Dz}^2 \right)}} \left(
\frac{m_{\Bp}^3}{m_{\Bz}^3} \frac{\tau_0}{\tau_+} \right) \right).\label{eq:C00}
\end{align}
We note that this result arises when both isospin amplitudes and topological
amplitudes are considered together. After getting the expression for $\Czz$ it
is pertinent that we do the required numerical analysis taking the current
experimental data into account.

\section{Numerical analysis}\label{sec:numerical}

\subsection{Experimental data}%
For $B \to D\Dbar$ decays, we have results from many
experiments~\cite{Aaij:2016yip, Aaij:2013fha, Rohrken:2012ta, Fratina:2007zk,
	Aubert:2008ah, Adachi:2008cj, Aubert:2006ia}.  But for consistency, we consider
the PDG~\cite{PDG} data in this paper and for comparison we take into account
the recent results on $C_{+-}$ by LHCb \cite{Aaij:2016yip} and by the heavy
flavor averaging group, HFLAV \cite{Amhis:2016xyh} as well. The experimental
data are,
\begin{subequations}\label{eq:data-1}
	\begin{align}
	\Bpm &= (2.11 \pm 0.18) \times 10^{-4},\\
	\Cpm &=
	\begin{cases}
	-0.22 \pm 0.24 & \text{(PDG)},\\ 0.26^{+0.18}_{-0.17}(\textrm{stat}) \pm 0.02
	(\textrm{syst}) & \text{(LHCb)},\\ -0.13 \pm 0.10 & \text{(HFLAV)},
	\end{cases}\\
	\Bzz &= (1.4 \pm 0.7) \times 10^{-5},\\
	\Bch &= (3.8 \pm 0.4) \times 10^{-4},\\
	\ACP &= -0.03 \pm 0.07.
	\end{align}
\end{subequations}
It is important to note that the errors in $\Bzz$, $\Cpm$ and $\ACP$ are large
enough to make them consistent with $0$ within $2\sigma$, and the \CP asymmetry
$\Czz$ is not yet measured experimentally. It must be noted that in the
averaging done by PDG for $C_{+-}$ both 2007 \cite{Fratina:2007zk} and 2012
\cite{Rohrken:2012ta} Belle results are considered, while the HFLAV averaging
for $C_{+-}$ considers only 2012 \cite{Rohrken:2012ta} Belle result. Since, the
2012 Belle result supersedes the 2007 Belle result (see
Ref.~\cite{Rohrken:2012ta}), we consider the HFLAV averaging of $C_{+-}$ to be
more reliable than the one done by PDG.

\subsection{Estimates of magnitudes of decay amplitudes}%

\begin{figure}[hbtp]
	\centering %
	\includegraphics[scale=1]{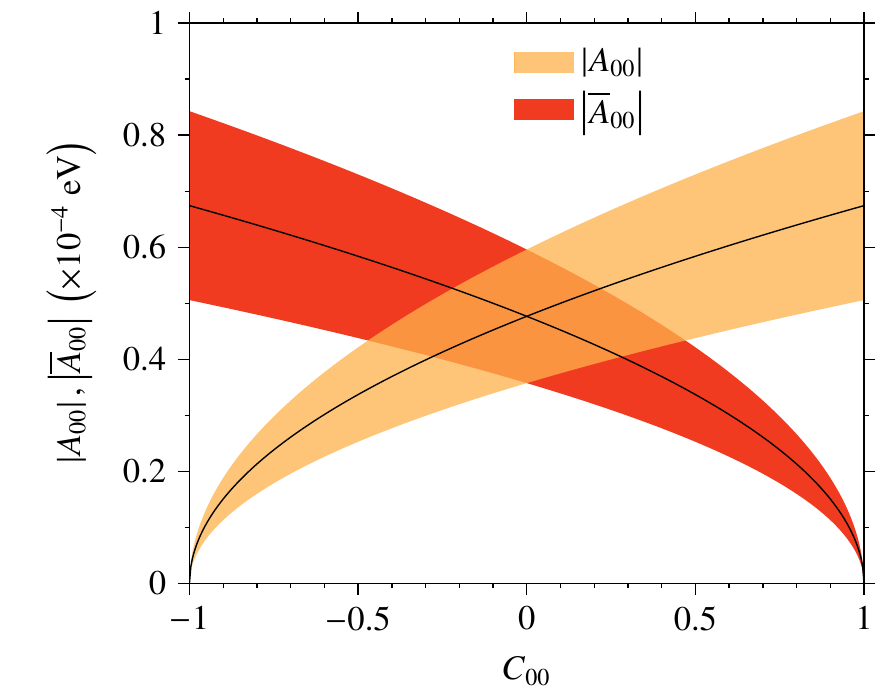}%
	\caption{Predictions of $\modulus{\Azz}$ and $\modulus{\Abzz}$ over the
		physically allowed range of $\Czz$. The colored band denotes the $1\sigma$
		error.} %
	\label{fig:Azz-Azzb-predictions}
\end{figure}

Using the experimental data from Eq.~\eqref{eq:data-1} in the expressions for
the moduli of the various amplitudes as given in Eq.~\eqref{eq:amp-mods} we get
the following estimates, with the errors combined in quadrature,
\begin{subequations}\label{eq:amp-mod-numerical}
	\begin{align}
	\modulus{\Apm} &=
	\begin{cases}
	\left( 1.637 \pm 0.261 \right)\times 10^{-4} \textrm{ eV} & \text{(PDG)},\\
	\left( 2.080 \pm 0.174 \right)\times 10^{-4} \textrm{ eV} & \text{(LHCb)},\\
	\left( 1.729 \pm 0.124 \right)\times 10^{-4} \textrm{ eV} & \text{(HFLAV)},
	\end{cases} \\
	\modulus{\Abpm} &=
	\begin{cases}
	\left( 2.047 \pm 0.261 \right)\times 10^{-4} \textrm{ eV} & \text{(PDG)},\\
	\left( 1.594 \pm 0.174 \right)\times 10^{-4} \textrm{ eV} & \text{(LHCb)},\\
	\left( 1.970 \pm 0.124 \right)\times 10^{-4} \textrm{ eV} & \text{(HFLAV)},
	\end{cases} \\
	\modulus{\Azm} &= \left( 2.358 \pm 0.150 \right)\times 10^{-4} \textrm{ eV}, \\
	\modulus{\Abpz} &= \left( 2.430 \pm 0.152 \right)\times 10^{-4} \textrm{ eV}.
	\end{align}
\end{subequations}
Since $\Czz$ is not yet known experimentally, we can only predict
$\modulus{\Azz}$ and $\modulus{\Abzz}$ in the physically allowed range of
$\Czz$. This is shown in Fig.~\ref{fig:Azz-Azzb-predictions}. We find that
\begin{equation}\label{eq:Azz-Abzz-up-limit}
0 \leqslant \modulus{\Azz}, \modulus{\Abzz} \leqslant \left(0.674 \pm
0.168\right)\times 10^{-4} \textrm{ eV}.
\end{equation}

\subsection{Numerical limits on magnitudes of isospin amplitudes}

\begin{figure}[hbtp]
\centering %
\includegraphics[width=\linewidth,keepaspectratio]{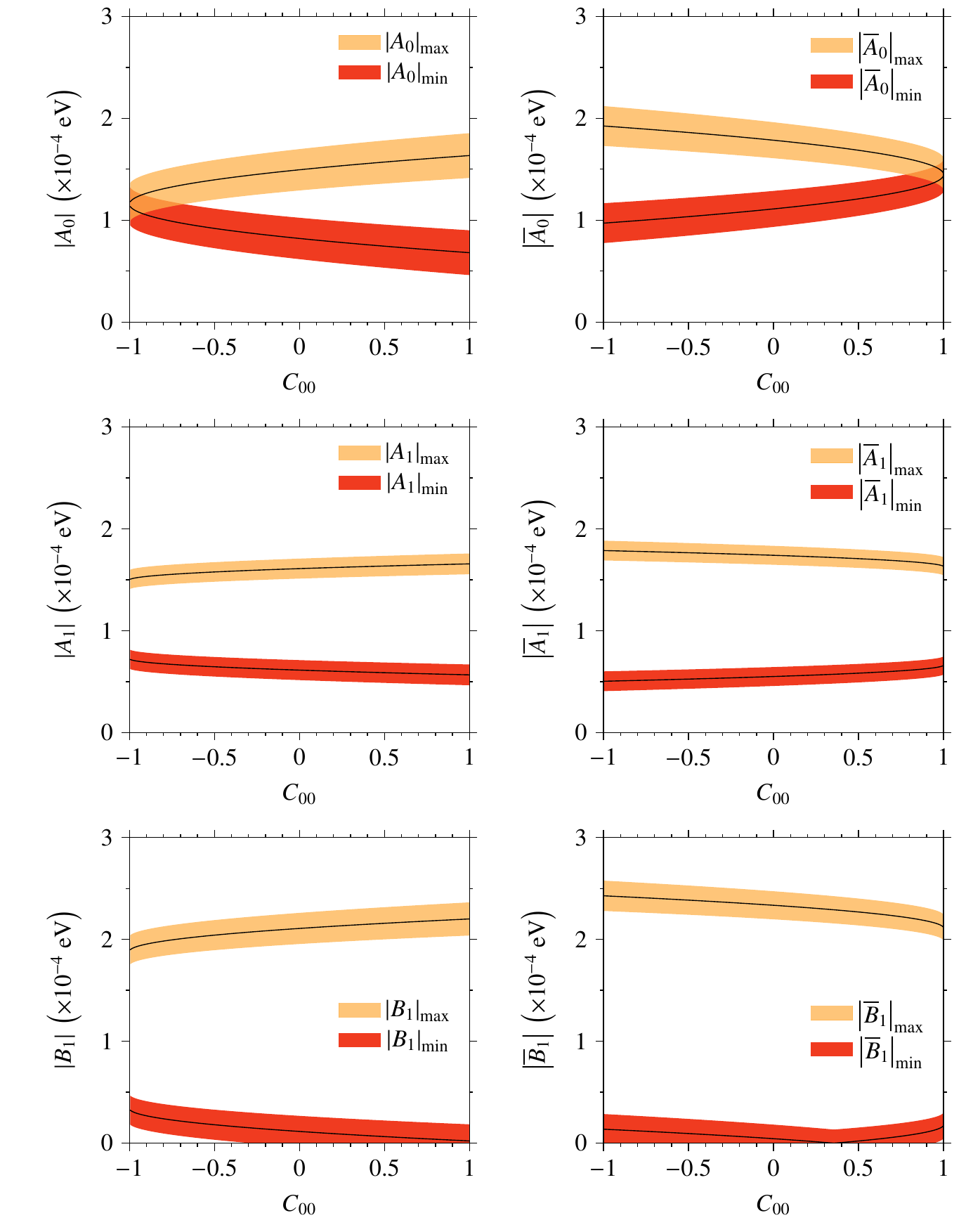} %
\caption{Comparison of upper and lower limits on $\modulus{A_0}$,
	$\modulus{\overline{A}_0}$, $\modulus{A_1}$, $\modulus{\overline{A}_1}$,
	$\modulus{B_1}$ and $\modulus{\overline{B}_1}$ assuming that $\Czz$ lies between
	$-1$ and $1$. Here we have used the PDG data~\cite{PDG} for $\Cpm$. It is easy
	to notice that the allowed ranges for $\modulus{B_1}$ and
	$\modulus{\overline{B}_1}$ are comparable with that of $\modulus{A_1}$ and
	$\modulus{\overline{A}_1}$ respectively.} %
\label{fig:B1-B1b-limits-PDG}
\end{figure}

\begin{figure}[hbtp]
\centering %
\includegraphics[width=\linewidth,keepaspectratio]{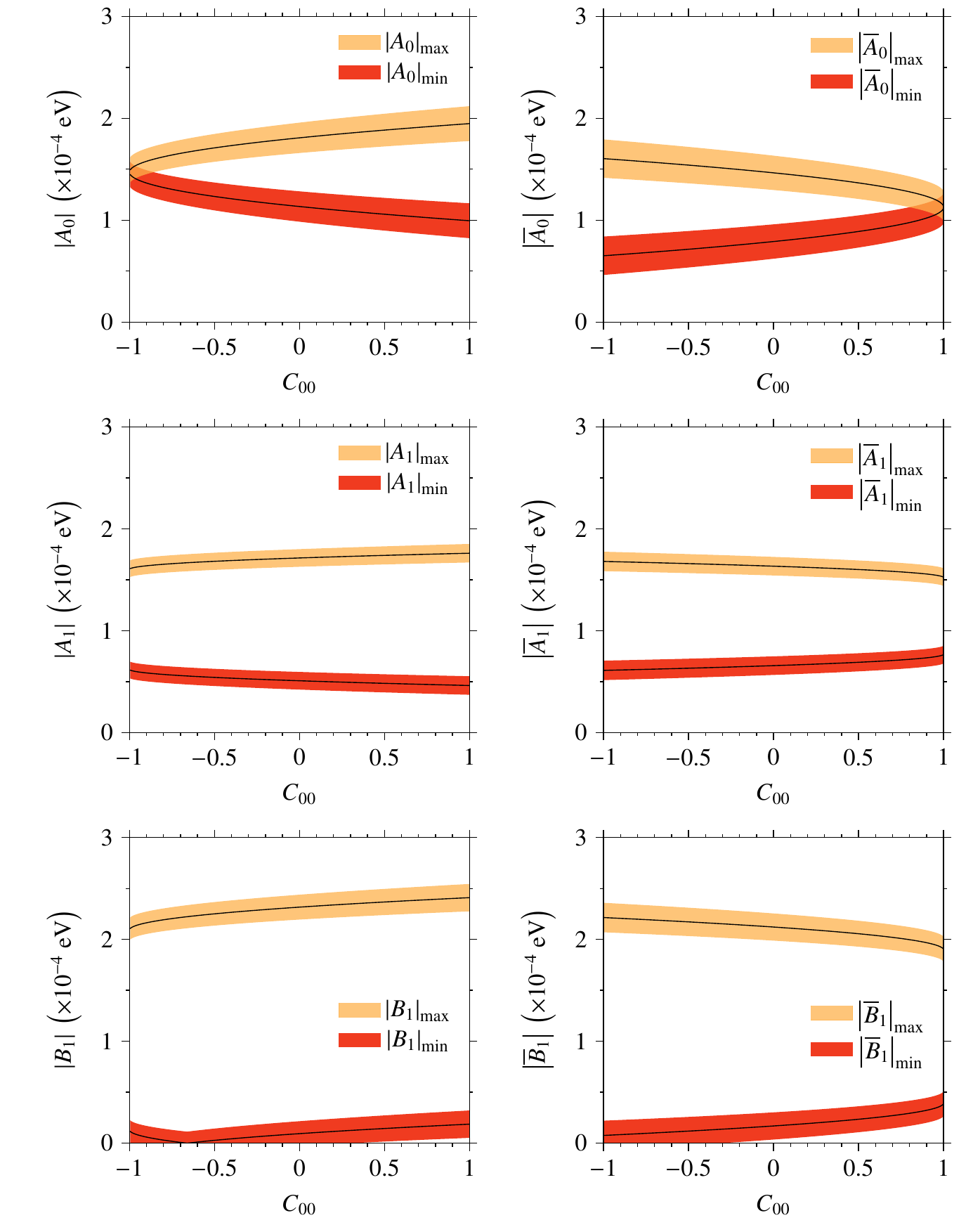} %
\caption{Same as Fig.~\ref{fig:B1-B1b-limits-PDG} except for using $C_{+-}$ as
	reported by the LHCb collaboration~\cite{Aaij:2016yip}.} %
\label{fig:B1-B1b-limits-LHCb}
\end{figure}

\begin{figure}[hbtp]
\centering %
\includegraphics[width=\linewidth,keepaspectratio]{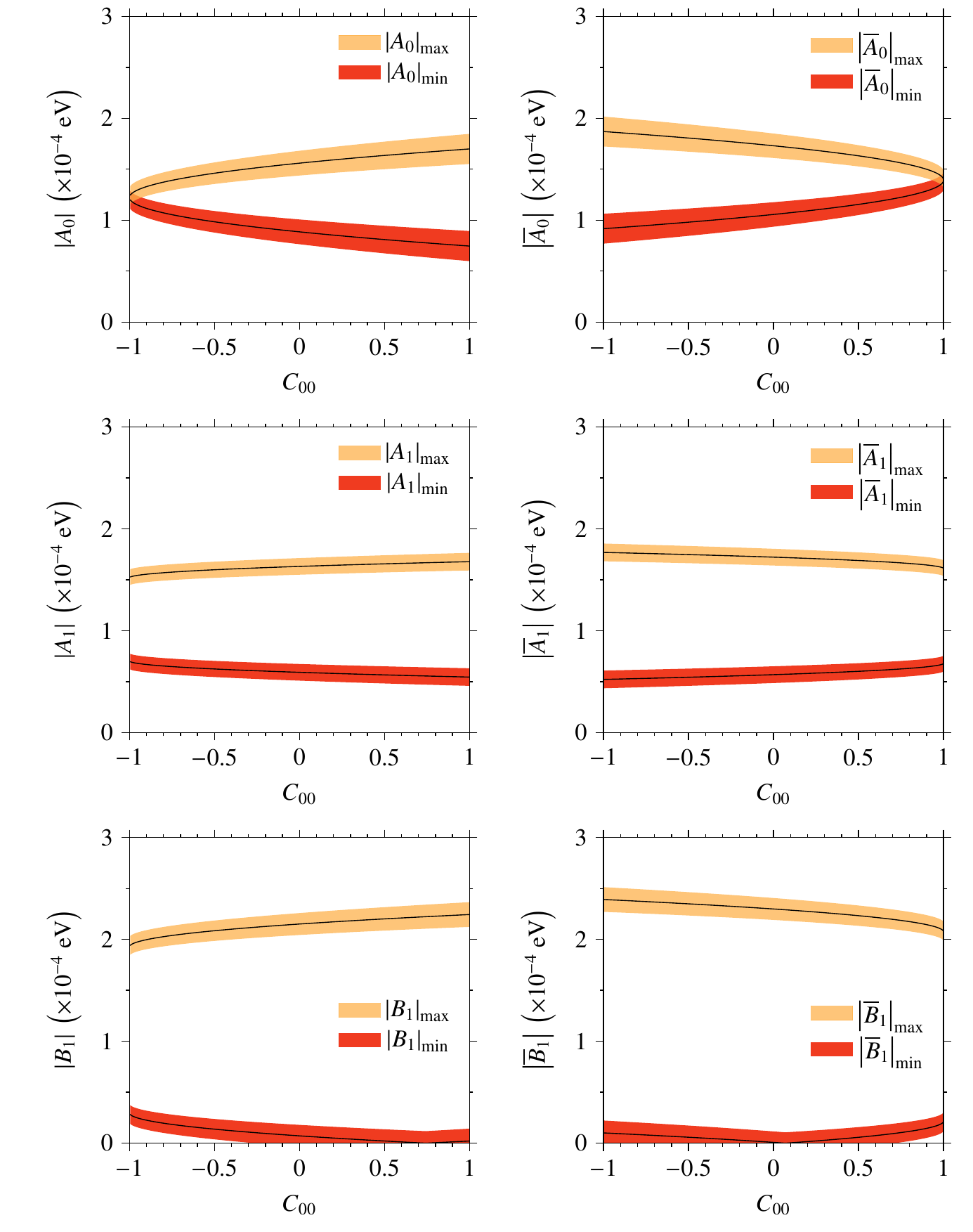}%
\caption{Same as Fig.~\ref{fig:B1-B1b-limits-PDG} except for using $C_{+-}$ as
	reported by the HFLAV~\cite{Amhis:2016xyh}.} %
\label{fig:B1-B1b-limits-HFLAV}
\end{figure}

The shaded regions in Figs.~\ref{fig:B1-B1b-limits-PDG},
\ref{fig:B1-B1b-limits-LHCb} and \ref{fig:B1-B1b-limits-HFLAV} show the allowed
maxima and minima of the magnitudes of isospin amplitudes: $\modulus{A_0}$,
$\modulus{\overline{A}_0}$, $\modulus{A_1}$, $\modulus{\overline{A}_1}$,
$\modulus{B_1}$ and $\modulus{\overline{B}_1}$, in the physically allowed range
of $\Czz$, and as permitted by the current experimental data taken from
PDG~\cite{PDG}, LHCb~\cite{Aaij:2016yip} and HFLAV~\cite{Amhis:2016xyh}
respectively. We have used Eqs.~\eqref{eq:bound-max-1} and
\eqref{eq:bound-max-2} to find out the maximum values of $\modulus{B_1}$ and
$\modulus{\overline{B}_1}$ respectively. For the minimum values we have three
cases for both Eq.~\eqref{eq:bound-min-1} and Eq.~\eqref{eq:bound-min-2} and we
have shown in the plots the absolute minimum out of the three minima
possibilities for both $\modulus{B_1}$ and $\modulus{\overline{B}_1}$. The
maxima and minima of $\modulus{A_1}$ and $\modulus{\overline{A}_1}$ were
evaluated in a similar manner using Eq.~\eqref{eq:A1-limits}. Finally, for the
maxima and minima of $\modulus{A_0}$ and $\modulus{\overline{A}_0}$, we have
made use of Eq.~\eqref{eq:A0-limits}.

\subsection{Predictions for \texorpdfstring{$\Czz$}{C00}}

In the absence of an experimental measurement, we can use Eq.~\eqref{eq:C00} to
predict a value for $\Czz$, which with current experimental data is
\begin{equation}\label{eq:Czz-prediction}
C_{00} =
\begin{cases}
4.081 \pm 4.530 & \textrm{(using PDG data)},\\%
-3.172 \pm 3.638 & \textrm{(using LHCb data)},\\%
2.721 \pm 2.699 & \textrm{(using HFLAV result)}.
\end{cases}
\end{equation}
Clearly, the central value of the predicted $\Czz$ lies outside the physically
allowed region for $\Czz$: $-1 \leqslant \Czz \leqslant 1$. However, the large
error in $\Czz$ essentially owes its origin to the fact that $\Bzz$, $\Cpm$ and
$\ACP$ are consistent with zero within $2\sigma$. Therefore, precise
measurements of $\Bzz$, $\Cpm$, $\ACP$ and an experimental determination of the
\CP asymmetry $\Czz$ would be very interesting.

\begin{figure}[hbtp]
	\centering%
	\includegraphics[scale=0.925]{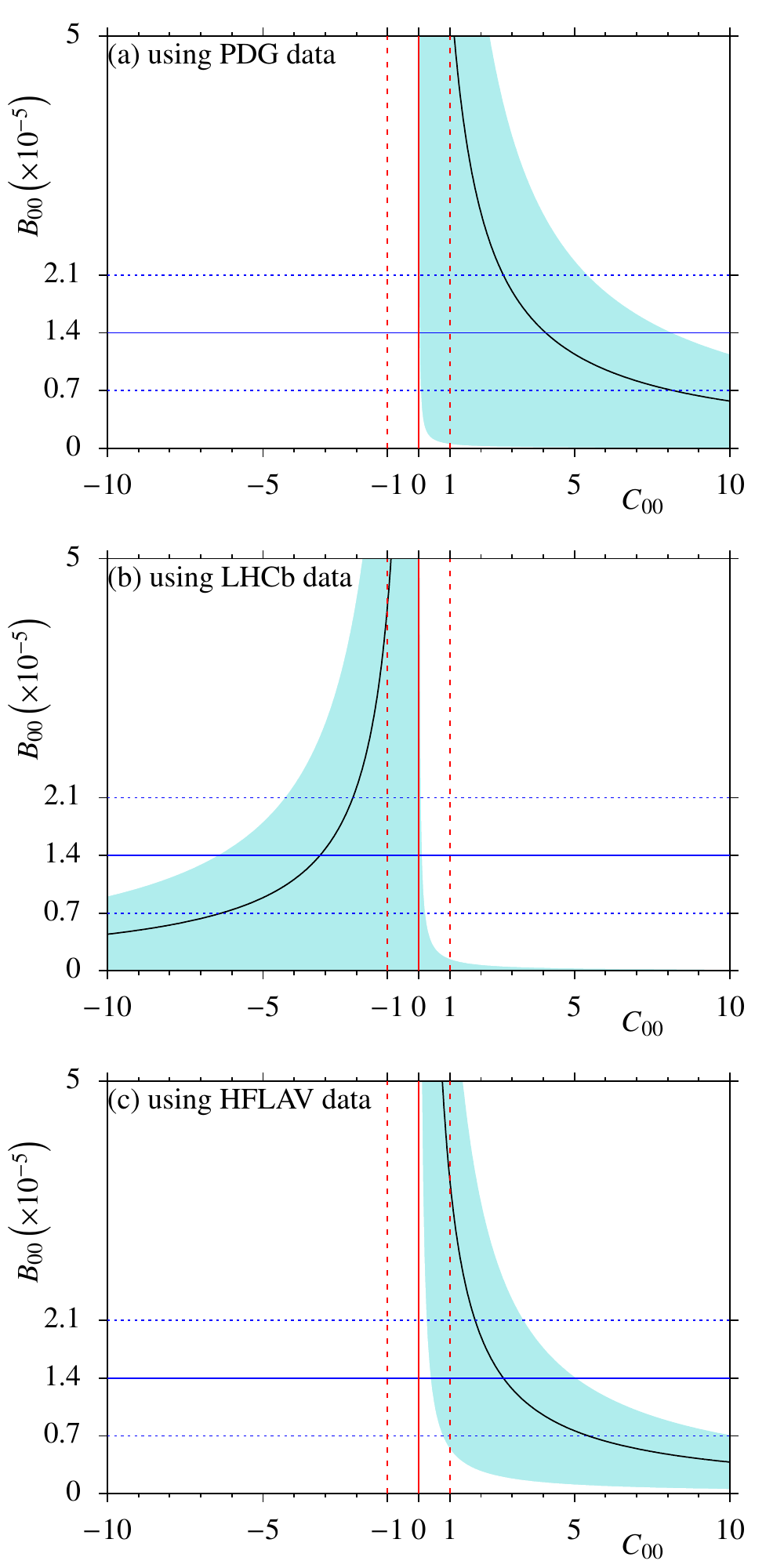}%
	\caption{Comparison of predicted value of $\Bzz$ from
		Eq.~\eqref{eq:B00} with the measured value of $\Bzz = (1.4 \pm 0.7) \times
		10^{-5}$, with $\Cpm$ taken from the PDG average, LHCb measurement and the HFLAV
		average. The shaded region shows the $1\sigma$ error on the predicted mean value
		of $\Bzz$ given by the black curve. The experimental value of $\Bzz$ lies within
		the blue dashed lines at $1\sigma$. Here for the purpose of illustration, we
		have included the unphysical regions of $\Czz$. The physically allowed region
		for $\Czz$ is within the red dashed lines.}%
	\label{fig:B00-comparison}
\end{figure}

From Eq.~\eqref{eq:C00}, one can write down the following expression for $\Bzz$:
\begin{equation}\label{eq:B00}
\Bzz = - \frac{\sqrt{\lambda\left(m_{\Bz}^2, m_{\Dz}^2, m_{\Dz}^2
		\right)}}{\Czz} \left( \frac{\Bpm \Cpm}{\displaystyle
	\sqrt{\lambda\left(m_{\Bz}^2, m_{\Dp}^2, m_{\Dp}^2 \right)}} + \frac{\Bch
	\ACP}{\displaystyle \sqrt{\lambda\left(m_{\Bp}^2, m_{\Dp}^2, m_{\Dz}^2 \right)}}
\left( \frac{m_{\Bp}^3}{m_{\Bz}^3} \frac{\tau_0}{\tau_+} \right) \right).
\end{equation}
Using Eq.~\eqref{eq:B00} we can predict the value of $\Bzz$ in the physically
allowed region of $\Czz$. In Fig.~\ref{fig:B00-comparison} we provide a
comparison of the predicted behaviour of $\Bzz$ with the experimental
measurement.

\subsection{Discussion on numerical analysis}

If we consider the mean values alone in Eq.~\eqref{eq:amp-mod-numerical} and the
upper limit from Eq.~\eqref{eq:Azz-Abzz-up-limit}, then we can arrange the
moduli of decay amplitudes in the following order,
\begin{equation*}
\modulus{\Azm} > \modulus{\Apm} > \modulus{\Azz}, ~\text{and}~ \modulus{\Abpz} >
\modulus{\Abpm} > \modulus{\Abzz},
\end{equation*}
consistent with the observation that $\Bch > \Bpm > \Bzz$. Moreover, considering
the mean values again we find that,
\begin{align*}
\modulus{\Apm} + \modulus{\Azz} < \modulus{\Azm} & & \text{(for PDG data)},\\%
\modulus{\Apm} + \modulus{\Azz} > \modulus{\Azm} & & \text{(for LHCb and
	HFLAV data)},\\%
\modulus{\Abpm} + \modulus{\Abzz} < \modulus{\Abpz} & & \text{(for LHCb
	data)},\\%
\modulus{\Abpm} + \modulus{\Abzz} > \modulus{\Abpz} & & \text{(for PDG and HFLAV
	data)}
\end{align*}
which lead to the possibilities $\modulus{B_1} \neq 0$, $\modulus{B_1} = 0$,
$\modulus{\overline{B}_1} \neq 0$ and $\modulus{\overline{B}_1} = 0$
respectively. However, if we consider the $1\sigma$ errors, both $\modulus{B_1}$
and $\modulus{\overline{B}_1}$ can vanish for all cases under our consideration.
This can be easily seen from Figs.~\ref{fig:B1-B1b-limits-PDG},
\ref{fig:B1-B1b-limits-LHCb} and \ref{fig:B1-B1b-limits-HFLAV}. From these
figures we also observe that the allowed ranges for $\modulus{A_1}$,
$\modulus{\overline{A}_1}$, $\modulus{B_1}$ and $\modulus{\overline{B}_1}$ are
all very similar. It must be noted that we have $9$ free parameters in our
formalism ($A'_0$, $A''_0$, $A'_1$, $A''_1$, $B'_1$, $B''_1$, $\delta_0$,
$\delta_1$, $\beta$) and currently we have experimental information about
$\sin\beta$ and $5$ out of the $6$ observables (viz., $\Bpm$, $\Bzz$, $\Bch$,
$\Cpm$, $\ACP$ and not yet $\Czz$). Therefore, it is not possible, at current,
to do a meaningful $\chi^2$-analysis and look for best fit values of the isospin
amplitudes in order to make a comparison. Even adding the observable $S_{+-}$
(and $S_{00}$ which is not yet measured) does not make any difference, since the
addition of these observables also leads to consideration of additional free
parameters. Nevertheless, as experimental data becomes available in the future
for all possible observables related to the decay modes under our consideration,
it would eventually be possible to do a meaningful $\chi^2$-analysis and study
the individual isospin amplitudes and strong phases in a clear manner.

As we have noted earlier the predicted value of $\Czz$ (see
Eq.~\eqref{eq:Czz-prediction}) has the mean value completely outside the
physically allowed region with large error which makes it consistent with zero
within about $1\sigma$. However, if we analyze Fig.~\ref{fig:B00-comparison} we
find that by looking in the window of observed range for $\Bzz$, within
$1\sigma$ standard deviation, the region $\Czz \geqslant 0$ is favoured by both
PDG and HFLAV data, while both positive and negative values of $\Czz$ are
allowed if we consider LHCb data alone. If we go to higher standard deviations,
the full physical range of $\Czz$ is allowed by the existing data, consistent
with Eq.~\eqref{eq:Czz-prediction}. It must be noted that, from
Eq.~\eqref{eq:B00} as well as from Fig.~\ref{fig:B00-comparison} it is clear
that the prediction for $\Bzz$ has a singularity at $\Czz=0$. Thus, if $\Czz$ is
experimentally measured to be non-zero, then we expect $\Bzz$ to have larger
value than the currently measured value which is consistent with zero at
$2\sigma$ level. This is by assuming that the other measurements, as given in
Eq.~\eqref{eq:data-1}, remain unchanged. A larger $\Bzz$ would imply significant
contribution from diagrams such as the $W$-exchange diagrams. Thus, this
interplay of $\Bzz$ and $\Czz$ measurements could lead to some potential search
for new physics.

We would like to emphasize that the large errors in
Eq.~\eqref{eq:Czz-prediction} can be reduced if we have precise measurements of
$\Bzz$, $\Cpm$ and $\ACP$ which are all currently consistent with zero within
$2\sigma$. To illustrate this point, let us consider a scenario in which future
experimental analyses with larger data sets give us values of $\Bzz$, $\Cpm$ and
$\ACP$ with their central values unchanged but with reduced errors. This
scenario is hypothetical because future experiments will not only shrink the
errors but also shift the central values in general. Nevertheless, to put our
emphasis on precise measurement of $\Bzz$, $\Cpm$ and $\ACP$ on a quantitative
basis, we can probe the prospect of Belle II experiment which is expected to
have 50 times larger integrated luminosity than Belle \cite{Belle:2017wmf}. In
such a scenario, we can na\"ively expect the errors on $\Bzz$, $\Cpm$ and $\ACP$
(taking the HFLAV averages as an example) to get scaled down by a factor of
roughly $1/\sqrt{50}$. Using our method, such reduced errors on the
above-mentioned observables will render an error of about $0.4$ for $C_{00}$. 
If the central value of $C_{00}$ is still significantly larger than 1 compared
to this new error, it will be an interesting hint of new physics at work. It is
also important to note that as $B^0 \to \DzDzb$ involves the $W$-exchange,
QCD-penguin annihilation and electroweak-penguin annihilation diagrams, a more
precise determination of observables related to this mode provides an ideal
means to probe the strong dynamics in these topological amplitudes.

\section{Conclusions}\label{sec:conclusion}

In this paper we have analyzed the $B \to D\Db$ decay modes in terms of isospin
amplitudes and the topological amplitudes. This leads us to predict the value of
the \CP asymmetry $\Czz$ in $\Bz \to \Dz\Dzb$ mode to be $4.081 \pm 4.530$, or
$-3.172 \pm 3.638$, or $2.721 \pm 2.699$ depending on whether we use the $\Cpm$
value as reported by PDG, or LHCb, or HFLAV, respectively. Though the central
values are all outside the physically allowed range for $\Czz$, the predictions
are consistent with zero due to large errors. The errors in $\Czz$ predictions
are large because of very large errors in $\Bzz$, $\Cpm$ and $\ACP$ all of which
enter the expression for $\Czz$. With more precise measurements of $\Bzz$,
$\Cpm$ and $\ACP$, and an experimental observation of $\Czz$ it would be
possible to make a better comparison of the observation with prediction using
Eq.~\eqref{eq:C00}. Further experimental results from the time-dependent decay
rates for the modes $\Bz \to \Dz\Dzb$ and $\Bz \to \Dp \Dm$ would pave the way
for a complete meaningful $\chi^2$-analysis which can be used to determine the
contributions of various isospin amplitudes, as well as strong phases that take
part in the $B \to D\Db$ decays under consideration. Furthermore, the
correlation between $B_{00}$ and $C_{00}$ is an interesting aspect that can be
probed in ongoing and future particle physics experiments such as LHCb and
Belle~II.

\acknowledgments

The works of HYC and CWC were supported in part by the Ministry of Science  and
Technology (MOST) of R.O.C.\ Grant Nos. 04-2112-M-001-022 and
104-2628-M-002-014-MY4 respectively. The work of CSK was supported by the NRF
grant funded by Korea government of the MEST (No. 2016R1D1A1A02936965). DS would
like to thank The Institute of Mathematical Sciences, Chennai, India, and
Institute of Physics, Academia Sinica, Taiwan, R.O.C.\ where some part of this
work was done, for hospitality.


\begin{thebibliography}{99}
	
\bibitem{Sakharov:1967dj} A.~D.~Sakharov, ``Violation of CP invariance, C
asymmetry, and baryon asymmetry of the universe,'' Pisma Zh.\ Eksp.\ Teor.\
Fiz.\  {\bf 5}, 32 (1967) [JETP Lett.\  {\bf 5}, 24 (1967)] [Sov.\ Phys.\ Usp.\
{\bf 34}, 392 (1991)] [Usp.\ Fiz.\ Nauk {\bf 161}, 61 (1991)].

\bibitem{Cabibbo:1963yz} N.~Cabibbo, ``Unitary Symmetry and Leptonic Decays,''
Phys.\ Rev.\ Lett.\  {\bf 10}, 531 (1963).

\bibitem{Kobayashi:1973fv} M.~Kobayashi and T.~Maskawa, ``CP Violation in the
Renormalizable Theory of Weak Interaction,'' Prog.\ Theor.\ Phys.\ {\bf 49}, 652
(1973).

\bibitem{Gavela:1993ts} M.~B.~Gavela, P.~Hernandez, J.~Orloff and O.~Pene,
``Standard model CP violation and baryon asymmetry,'' Mod.\ Phys.\ Lett.\ A {\bf
	9}, 795 (1994).

\bibitem{Gavela:1994dt} M.~B.~Gavela, P.~Hernandez, J.~Orloff, O.~Pene and
C.~Quimbay, ``Standard model CP violation and baryon asymmetry. Part 2: Finite
temperature,'' Nucl.\ Phys.\ B {\bf 430}, 382 (1994)
doi:10.1016/0550-3213(94)00410-2 [hep-ph/9406289].

\bibitem{Huet:1994jb} P.~Huet and E.~Sather, ``Electroweak baryogenesis and
standard model CP violation,'' Phys.\ Rev.\ D {\bf 51}, 379 (1995).

\bibitem{Antonelli:2009ws} M.~Antonelli {\it et al.}, ``Flavor Physics in the
Quark Sector,'' Phys.\ Rept.\  {\bf 494}, 197 (2010).

\bibitem{Hocker:2006xb} A.~Hocker and Z.~Ligeti, ``CP violation and the CKM
matrix,'' Ann.\ Rev.\ Nucl.\ Part.\ Sci.\  {\bf 56}, 501 (2006).

\bibitem{Artuso:2015swg} M.~Artuso, G.~Borissov and A.~Lenz, ``CP violation in
the $B_s^0$ system,'' Rev.\ Mod.\ Phys.\  {\bf 88}, no. 4, 045002 (2016).

\bibitem{Gershon:2016fda} T.~Gershon and V.~V.~Gligorov, ``$CP$ violation in the
$B$ system,'' Rept.\ Prog.\ Phys.\  {\bf 80}, no. 4, 046201 (2017).

\bibitem{Bel:2015wha} L.~Bel, K.~De Bruyn, R.~Fleischer, M.~Mulder and
N.~Tuning, ``Anatomy of $ B\to D\overline{D} $ decays,'' JHEP {\bf 1507}, 108
(2015).

\bibitem{Jung:2014jfa} M.~Jung and S.~Schacht, ``Standard model predictions and
new physics sensitivity in $B \to DD$ decays,'' Phys.\ Rev.\ D {\bf 91}, no. 3,
034027 (2015).

\bibitem{Mohammadi:2011zz} B.~Mohammadi and H.~Mehraban, ``Final state
interaction in $B^0 \to D^0 \overline{D}^0$,'' JHEP {\bf 1107}, 089 (2011).

\bibitem{Lu:2010gg} L.~X.~Lu, Z.~J.~Xiao, S.~W.~Wang and W.~J.~Li, ``The Double
charm decays of B Mesons in the mSUGRA model,'' Commun.\ Theor.\ Phys.\  {\bf
56}, 125 (2011).

\bibitem{Li:2009xf} R.~H.~Li, X.~X.~Wang, A.~I.~Sanda and C.~D.~Lu, ``Decays of
$B$ meson to two charmed mesons,'' Phys.\ Rev.\ D {\bf 81}, 034006 (2010).

\bibitem{Kim:2008ex} C.~S.~Kim, R.~M.~Wang and Y.~D.~Yang, ``Studying Double
Charm Decays of $B_{u,d}$ and $B_{s}$ Mesons in the MSSM with $R$-parity
Violation,'' Phys.\ Rev.\ D {\bf 79}, 055004 (2009).

\bibitem{Gronau:2008ed} M.~Gronau, J.~L.~Rosner and D.~Pirjol, ``Small amplitude
effects in $B^0 \to D^{+} D^{-}$ and related decays,'' Phys.\ Rev.\ D {\bf 78},
033011 (2008).

\bibitem{Li:2007rk} Y.~Li and J.~Hua, ``Study of pure annihilation decays
$B_{d,s} \to D^0 \overline{D}^0$,'' Chin.\ Phys.\ C {\bf 32}, 781 (2008).

\bibitem{Fleischer:2007zn} R.~Fleischer, ``Exploring $CP$ violation and penguin
effects through $B^0_{d} \to D^{+} D^{-}$ and $B^0_{s} \to D^+_{s} D^-_{s}$,''
Eur.\ Phys.\ J.\ C {\bf 51}, 849 (2007).

\bibitem{Chen:2005rp} C.~H.~Chen, C.~Q.~Geng and Z.~T.~Wei, ``Factorization and
polarization in two charmed-meson $B$ decays,'' Eur.\ Phys.\ J.\ C {\bf 46}, 367
(2006).

\bibitem{Datta:2003va} A.~Datta and D.~London, ``Extracting $\gamma$ from
$B^0_d(t) \to D^{(*)+} D^{(*)-}$ and $B^0_d \to D^{(*)+}_s D^{(*)-}$ decays,''
Phys.\ Lett.\ B {\bf 584}, 81 (2004).

\bibitem{Xing:1999yx} Z.~z.~Xing, ``$CP$ violation in $B_d \to D^+ D^-, D^{*+}
D^-, D^+ D^{*-}$ and $D^{*+} D^{*-}$ decays,'' Phys.\ Rev.\ D {\bf 61}, 014010
(1999).

\bibitem{Pham:1999fy} X.~Y.~Pham and Z.~z.~Xing, ``$CP$ asymmetries in $B_d \to
D^{*+} D ^{*-}$ and $B_s \to D^{*+}_s D^{*-}_s$: $P$ wave dilution, penguin and
rescattering effects,'' Phys.\ Lett.\ B {\bf 458}, 375 (1999).

\bibitem{Fleischer:1999nz} R.~Fleischer, ``Extracting $\gamma$ from $B_{s(d)}
\to J/\psi K_{S}$ and $B_{d(s)} \to D^+_{d(s)} D^-_{d(s)}$,'' Eur.\ Phys.\ J.\ C
{\bf 10}, 299 (1999).

\bibitem{Xing:1998ca} Z.~z.~Xing, ``Measuring $CP$ violation and testing
factorization in $ B_d \to D^{*\pm}~D^{\mp}$ and $B_s \to D^{*\pm}_s~D^{\mp}_s $
decays,'' Phys.\ Lett.\ B {\bf 443}, 365 (1998).

\bibitem{Sanda:1997pm} A.~I.~Sanda and Z.~z.~Xing, ``Towards determining
$\phi_1$ with $B \to D^{(*)} \bar D^{(*)}$,'' Phys.\ Rev.\ D {\bf 56}, 341
(1997).

\bibitem{Gronau:1995hm} M.~Gronau, O.~F.~Hernandez, D.~London and J.~L.~Rosner,
``Decays of $B$ mesons to two pseudoscalars in broken $SU(3)$ symmetry,'' Phys.\
Rev.\ D {\bf 52}, 6356 (1995).

\bibitem{Kramer:1994in} G.~Kramer, W.~F.~Palmer and H.~Simma, ``$CP$ violation
and strong phases from penguins in $B^{\pm} \to PP$ and $B^{\pm} \to VP$
decays,'' Z.\ Phys.\ C {\bf 66}, 429 (1995).

\bibitem{Aleksan:1993qk} R.~Aleksan, A.~Le Yaouanc, L.~Oliver, O.~Pene and
J.~C.~Raynal, ``The Decay $B \to D \overline{D}^* + D^* \overline{D}$ in the
heavy quark limit and tests of $CP$ violation,'' Phys.\ Lett.\ B {\bf 317}, 173
(1993).

\bibitem{Zeppenfeld:1980ex} D.~Zeppenfeld, ``$SU(3)$ Relations for $B$ Meson
Decays,'' Z.\ Phys.\ C {\bf 8}, 77 (1981).

\bibitem{Chau:1982da} L.~L.~Chau, ``Quark Mixing in Weak Interactions,'' Phys.\
Rept.\  {\bf 95}, 1 (1983).

\bibitem{Chau:1986du} L.~L.~Chau and H.~Y.~Cheng, ``Quark Diagram Analysis of
Two-body Charm Decays,'' Phys.\ Rev.\ Lett.\  {\bf 56}, 1655 (1986).

\bibitem{Chau:1987tk} L.~L.~Chau and H.~Y.~Cheng, ``Analysis of Exclusive
Two-Body Decays of Charm Mesons Using the Quark Diagram Scheme,'' Phys.\ Rev.\ D
{\bf 36}, 137 (1987).

\bibitem{Chau:1990ay} L.~L.~Chau, H.~Y.~Cheng, W.~K.~Sze, H.~Yao and B.~Tseng,
``Charmless nonleptonic rare decays of $B$ mesons,'' Phys.\ Rev.\ D {\bf 43},
2176 (1991) Erratum: [Phys.\ Rev.\ D {\bf 58}, 019902 (1998)].

\bibitem{Kim:1998sh} C.~S.~Kim, D.~London and T.~Yoshikawa, ``Using $B_s^0$
decays to determine the $CP$ angles $\alpha$ and $\gamma$,'' Phys.\ Rev.\ D {\bf
57}, 4010 (1998).

\bibitem{Chiang:2003jn}
  C.~W.~Chiang and J.~L.~Rosner,
  ``New physics contributions to the $B \to \phi K_S$ decay,''
  Phys.\ Rev.\ D {\bf 68}, 014007 (2003).

\bibitem{Chiang:2003rb}
  C.~W.~Chiang, M.~Gronau and J.~L.~Rosner,
  ``Two body charmless B decays involving eta and eta-prime,''
  Phys.\ Rev.\ D {\bf 68}, 074012 (2003).

\bibitem{Chiang:2003pm}
  C.~W.~Chiang, M.~Gronau, Z.~Luo, J.~L.~Rosner and D.~A.~Suprun,
  ``Charmless $B \to V P$ decays using flavor SU(3) symmetry,''
  Phys.\ Rev.\ D {\bf 69}, 034001 (2004).

\bibitem{Chiang:2004nm}
  C.~W.~Chiang, M.~Gronau, J.~L.~Rosner and D.~A.~Suprun,
  ``Charmless $B \to PP$ decays using flavor SU(3) symmetry,''
  Phys.\ Rev.\ D {\bf 70}, 034020 (2004).

\bibitem{Chiang:2006ih}
  C.~W.~Chiang and Y.~F.~Zhou,
  ``Flavor SU(3) analysis of charmless $B$ meson decays to two pseudoscalar mesons,''
  JHEP {\bf 0612}, 027 (2006).

\bibitem{Chiang:2007qh}
  C.~W.~Chiang and Y.~F.~Zhou,
  ``Flavor SU(3) analysis of charmless $B \to PP$ decays,''
  J.\ Phys.\ Conf.\ Ser.\  {\bf 110}, 052056 (2008).

\bibitem{Chiang:2008vc}
  C.~W.~Chiang, M.~Gronau and J.~L.~Rosner,
  ``Examination of Flavor SU(3) in $B$, $B_s \to K \pi$ Decays,''
  Phys.\ Lett.\ B {\bf 664}, 169 (2008).

\bibitem{Cheng:2010ry}
  H.~Y.~Cheng and C.~W.~Chiang,
  ``Two-body hadronic charmed meson decays,''
  Phys.\ Rev.\ D {\bf 81}, 074021 (2010)

\bibitem{Cheng:2010vk}
  H.~Y.~Cheng and C.~W.~Chiang,
  ``Hadronic D decays involving even-parity light mesons,''
  Phys.\ Rev.\ D {\bf 81}, 074031 (2010).

\bibitem{Cheng:2011qh} H.~Y.~Cheng and S.~Oh, ``Flavor $SU(3)$ symmetry and QCD
factorization in $B \to PP$ and $PV$ decays,'' JHEP {\bf 1109}, 024 (2011).

\bibitem{Cheng:2012wr}
  H.~Y.~Cheng and C.~W.~Chiang,
  ``Direct CP violation in two-body hadronic charmed meson decays,''
  Phys.\ Rev.\ D {\bf 85}, 034036 (2012)
  Erratum: [Phys.\ Rev.\ D {\bf 85}, 079903 (2012)].

\bibitem{Cheng:2012xb}
  H.~Y.~Cheng and C.~W.~Chiang,
  ``SU(3) symmetry breaking and CP violation in $D \to PP$ decays,''
  Phys.\ Rev.\ D {\bf 86}, 014014 (2012).

\bibitem{Cheng:2016ejf}
  H.~Y.~Cheng, C.~W.~Chiang and A.~L.~Kuo,
  ``Global analysis of two-body $D\to VP$ decays within the framework of flavor symmetry,''
  Phys.\ Rev.\ D {\bf 93}, no. 11, 114010 (2016).

\bibitem{Aaij:2016yip} R.~Aaij {\it et al.} [LHCb Collaboration], ``Measurement
of $CP$ violation in $B^0 \!\rightarrow D^+ D^-$ decays,'' Phys.\ Rev.\ Lett.\
{\bf 117}, no. 26, 261801 (2016).

\bibitem{Aaij:2013fha} R.~Aaij {\it et al.} [LHCb Collaboration], ``First
observations of $\bar{B}_s^0 \to D^+D^-$, $D_s^+D^-$ and $D^0\bar{D}^0$
decays,'' Phys.\ Rev.\ D {\bf 87}, no. 9, 092007 (2013).

\bibitem{Rohrken:2012ta} M.~Rohrken {\it et al.} [Belle Collaboration],
``Measurements of Branching Fractions and Time-dependent CP Violating
Asymmetries in $B^{0} \to D^{(*)\pm}D^{\mp}$ Decays,'' Phys.\ Rev.\ D {\bf 85},
091106 (2012).

\bibitem{Fratina:2007zk} S.~Fratina {\it et al.} [Belle Collaboration],
``Evidence for CP violation in $B^0 \to D^+ D^-$ decays,'' Phys.\ Rev.\ Lett.\
{\bf 98}, 221802 (2007).

\bibitem{Aubert:2008ah} B.~Aubert {\it et al.} [BaBar Collaboration],
``Measurements of time-dependent CP asymmetries in $B^0 \to D^{(*)} + D^{(*)}$ -
decays,'' Phys.\ Rev.\ D {\bf 79}, 032002 (2009).

\bibitem{Adachi:2008cj} I.~Adachi {\it et al.} [Belle Collaboration],
``Measurement of the branching fraction and charge asymmetry of the decay $B^+
\to D^+ \bar D^0$ and search for $B^0 \to D^0 \bar D^0$,'' Phys.\ Rev.\ D {\bf
77}, 091101 (2008).

\bibitem{Aubert:2006ia} B.~Aubert {\it et al.} [BaBar Collaboration],
``Measurement of branching fractions and $CP$-violating charge asymmetries for
$B$-meson decays to $D^{(*)} \bar D^{(*)}$, and implications for the
Cabibbo-Kobayashi-Maskawa angle $\gamma$,'' Phys.\ Rev.\ D {\bf 73}, 112004
(2006).

\bibitem{PDG}  C. Patrignani \textit{et al.} (Particle Data Group), Chin. Phys.
C, \textbf{40}, 100001 (2016) and 2017 update.

\bibitem{Amhis:2016xyh} Y.~Amhis {\it et al.}, ``Averages of $b$-hadron,
$c$-hadron, and $\tau$-lepton properties as of summer 2016,'' arXiv:1612.07233
[hep-ex].

\bibitem{Belle:2017wmf} L.~Li Gioi [Belle and Belle II Collaborations], ``Belle
achievements and Belle II prospects for CP violation,'' J.\ Phys.\ Conf.\ Ser.\
{\bf 873}, no. 1, 012022 (2017).

\end{thebibliography}
\end{document}